\documentclass[11pt]{article}

\usepackage{graphicx,amsmath,amssymb,a4wide}

\begin{document}

\begin{center}
\begin{Large}
\textbf{Some comments on pinwheel tilings\\[0.5ex] and their 
diffraction}\vspace{3ex}
\end{Large}

\begin{large}
Uwe Grimm$^1$ and Xinghua Deng$^2$\vspace{1ex}
\end{large}
\end{center}

\begin{quote}
\noindent $^1$ Department of Mathematics and Statistics, The Open University, 
Walton Hall,\newline
\hphantom{$^1$} Milton Keynes MK7~6AA, UK

\noindent $^2$ University of Waterloo, 200 University Avenue West, Waterloo, 
Ontario, Canada\newline
\hphantom{$^2$} N2L 3G1\vspace{1.5ex}
\end{quote}

\begin{quote}
The pinwheel tiling is the paradigm for a substitution tiling with
circular symmetry, in the sense that the corresponding autocorrelation
is circularly symmetric. As a consequence, its diffraction measure is
also circularly symmetric, so the pinwheel diffraction consists of
sharp rings and, possibly, a continuous component with circular
symmetry. We consider some combinatorial properties of the tiles and
their orientations, and a numerical approach to the diffraction of weighted
pinwheel point sets.\vspace{1.5ex}
\end{quote}

\section{Introduction}

Diffraction is an essential tool for the determination of the atomic
structure of solids. The discovery of aperiodically ordered materials
in the early 1980s \cite{SBGC84,INF85} added a new dimension of
complexity to the inverse problem of structure determination
\cite{BG09,BG10}, and led to a renewed interest in mathematical
diffraction theory, which focuses on kinematic diffraction
\cite{RIMS,BG}. It also posed the question of what type of ordered
structures are possible, and how one would be able to detect
them. Regular model sets, which are structures derived from a
(periodic) lattice in a higher-dimensional space, show pure point
diffraction; see \cite{L08} and references therein. However, there are
many interesting systems that are not model sets, such as tilings with
circular symmetry; we refer to \cite{tao} for general background on
the theory of aperiodic order.

Pinwheel tilings of the plane, first introduced by Conway and Radin
\cite{R94,R95,R97,R99}, are arguably the simplest examples of
substitution tilings with circular symmetry. They are built entirely
of triangular tiles of edge lengths $1$, $2$ and $\sqrt{5}$. The
corresponding right triangles occur in both possible orientations, or
chiralities, so in fact a pinwheel tiling comprises two tile shapes
which are mirror images of each other.

\begin{figure}[t]
\centerline{\includegraphics[width=0.3\textwidth]{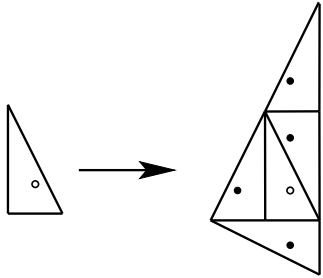}\vspace*{-2ex}}
\caption{\label{fig:infrule} Inflation rule of the pinwheel
  tiling. Control points are denoted by dots, with the open dot
  specifying the origin. The inflation rule is compatible with
  reflection, in the sense that the triangle of different chirality is
  replaced by the mirror image of the patch shown here.}
\end{figure}

A pinwheel tiling of the plane can be constructed as the fixed point
of the inflation rule of Figure~\ref{fig:infrule}. It consists of an
inflation with linear scaling factor $\sqrt{5}$ and a rotation by an
angle $\varphi=-\arctan(\frac{1}{2})\approx -26.565^{\circ}$, followed
by a dissection into five triangles of the original size and shape,
two of the same and three of the opposite chirality. The rotation
ensures that the original triangle reappears in the inflated patch,
and the sequence of iterates therefore converges to a space-filling
tiling of the plane. At the same time, because $\varphi$ is not a
rational multiple of $\pi$, it produces an irrational rotation, which
means that in each inflation step one obtains triangles in a new
direction.  In the limit, the set of available directions becomes
dense on the unit circle, and this produces the circular symmetry of
the structure; a proof of the circular symmetry of the autocorrelation
can be found in \cite{R97,MPS06}; see also \cite{F08} for
generalisations.

\section{Pinwheel control point sets}

Also shown in Figure~\ref{fig:infrule} are control points, which form
a point set $\varLambda$ that is mutually locally derivable (MLD) with
the pinwheel tiling, in the sense of \cite{BSJ91}. The control point
of a triangle is located at position $(\frac{1}{2},\frac{1}{2})$ with
respect to the right angle corner of the triangle. In other words, the
triangle with control point $(0,0)$ and its short edge along the
horizontal axis has vertices $(-\frac{1}{2},-\frac{1}{2})$,
$(\frac{1}{2},-\frac{1}{2})$ and $(-\frac{1}{2},\frac{3}{2})$, or
$(-\frac{1}{2},-\frac{1}{2})$, $(\frac{1}{2},-\frac{1}{2})$ and
$(\frac{1}{2},\frac{3}{2})$, depending on its chirality. Given a
control point $x\in\mathbb{R}^{2}\simeq \mathbb{C}$, a triangle can be
specified by a triple $(x,\omega,\chi)$, where $x\in\mathbb{C}$
denotes the corresponding control point, $\omega$ denotes the angle of
the short edge with respect to the horizontal axis and $\chi\in\{\pm
1\}$ specifies the chirality of the triangle.

\begin{figure}[t]
\centerline{\includegraphics[width=0.8\textwidth]{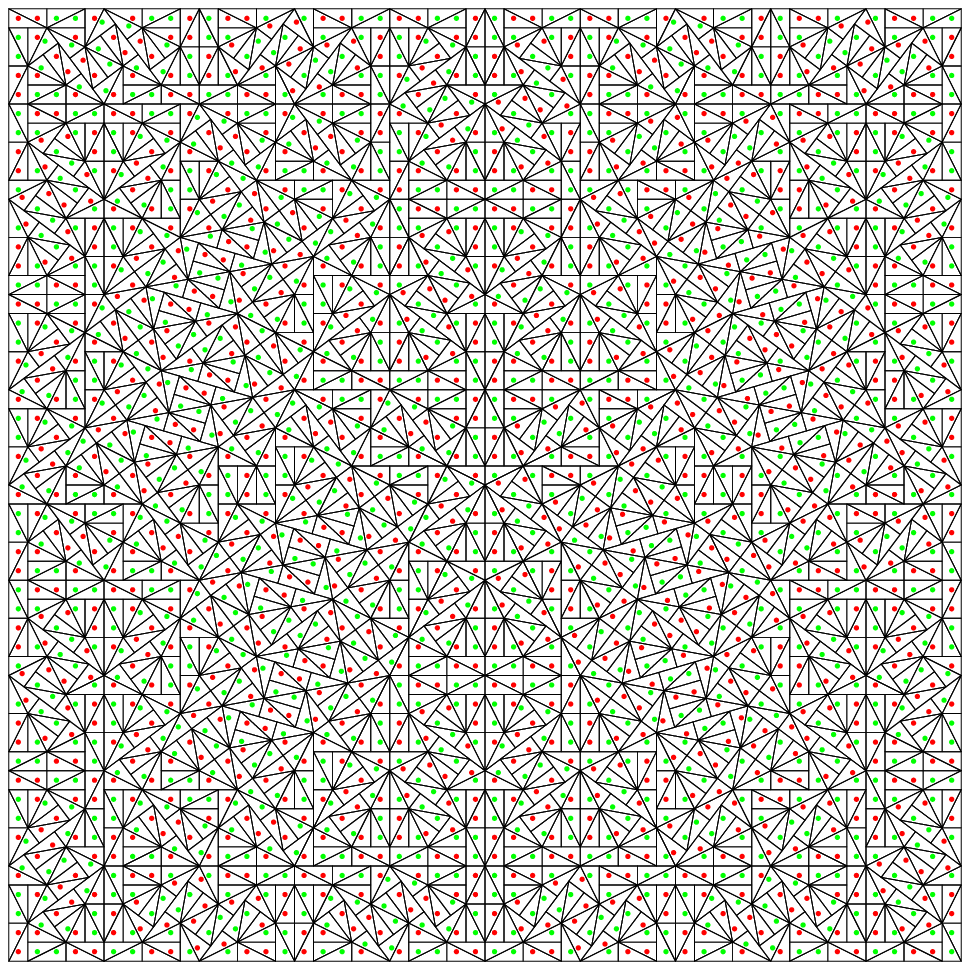}\vspace*{-2ex}}
\caption{\label{fig:pintwo} A patch of the pinwheel tiling. The
  control point in $\varLambda$ are coloured according to the two
  chiralities of the triangles.}
\end{figure}

The pinwheel point set $\varLambda$ forms a substitutive point set
with substitution rule
\begin{eqnarray}
\sigma(x,\omega,\chi)=\left\{
\begin{array}{lll}
(\sqrt{5}R_{\varphi}x, &\omega+\varphi-\chi\varphi,&\chi)\\
(\sqrt{5}R_{\varphi}x+R_{(\omega+\varphi-\chi\varphi+\frac{\chi\pi}{2})}, &
\omega+\varphi-\chi\varphi+\pi,&\chi)\\
(\sqrt{5}R_{\varphi}x+2R_{(\omega+\varphi-\chi\varphi+\frac{\chi\pi}{2})}, &
\omega+\varphi-\chi\varphi+\pi,&-\chi)\\
(\sqrt{5}R_{\varphi}x+R_{(\omega+\varphi-\chi\varphi+\pi)}, 
&\omega+\varphi-\chi\varphi+\pi,&-\chi)\\
(\sqrt{5}R_{\varphi}x+R_{(\omega+\varphi-\chi\varphi-\frac{\chi\pi}{2})}, 
&\omega+\varphi-\chi\varphi-\frac{\chi\pi}{2},&-\chi),\\
\end{array}
\right.
\end{eqnarray}
where $R_{\theta}=e^{i\theta}$ for $\theta\in\mathbb{R}$, so
multiplication by $R_\theta$ corresponds to a rotation in the complex
plane with the angle $\theta$. A patch of the pinwheel tiling, with
control points coloured according to the two chiralities of the
triangular tiles, is shown in Figure~\ref{fig:pintwo}.

The inflation rule $\sigma$ gives rise to the substitution matrix
\begin{eqnarray}
    M & = & \begin{pmatrix}
    e^{it0}+e^{it\pi} & 
    2e^{it(2\varphi-\pi)}+e^{it(2\varphi+\frac{\pi}{2})} \\
    2 e^{it\pi} + e^{it(-\frac{\pi}{2})} & e^{it(2\varphi)}+e^{it(2\varphi-\pi)}
   \end{pmatrix}\nonumber \\ & = & 
   \begin{pmatrix}
    1+y^{2} &  2xy^{-1}+xy \\
    2 y^{2} + 1/y^{-1} & x+xy^{-2}
    \end{pmatrix} 
\end{eqnarray}
where we set $x=e^{2i\varphi t}$ and $y=e^{i\pi t/2}$. This matrix was
used in \cite{MPS06} to derive the circular symmetry of the
autocorrelation. If we only wish to distinguish orientations modulo 90
degrees, we can choose $t=4$, so $y=1$, and the matrix simplifies to
\begin{equation}
    M = \begin{pmatrix}
    2 & 3x \\
    3 & 2x
   \end{pmatrix}.
\end{equation}
This matrix can be used to derive information about the subsets of
control points that belong to the same direction, in the sense that
they belong to the same power of $x$.

\section{Hierarchies of subsets of different orientation}

Let us consider what happens when we start from an initial patch
containing one triangle of each chirality with horizontal orientation,
and perform $n$ inflation steps. Denoting by $f^{\pm}_{n}(k)$ the
number of triangles of each chirality at step $n$ with orientation
factor $x^{k}$, where $0\le k\le n$, we find
\begin{equation}
    \begin{pmatrix}
    \sum_{k=0}^{n}f^{+}_{n}(k)\, x^{k}\\
    \sum_{k=0}^{n}f^{-}_{n}(k)\, x^{k}
    \end{pmatrix}
    \; = \; M^{n} \,\begin{pmatrix} 1 \\ 1 \end{pmatrix}.
\end{equation}
Multiplication by $M$ gives the following recursion for the coefficients
\begin{subequations}
\begin{eqnarray}
f^{+}_{n+1}(k) & = & 2\, f^{+}_{n}(k) + 3\, f^{-}_{n}(k-1)\, , \\
f^{-}_{n+1}(k) & = & 2\, f^{+}_{n}(k-1) + 3\, f^{-}_{n}(k)\, ,
\end{eqnarray}
\end{subequations}
for $0\le k\le n+1$,
where the appropriate initial conditions are $f^{\pm}_{0}(0)=1$
and $f^{\pm}_{n}(k)=0$ for all $k<0$ and $k>n$. 

Clearly, these numbers satisfy the relation
$f^{+}_{n}(k)=f^{-}_{n}(n-k)$ for all $0\le k\le n$, as can be
seen from the relation
\begin{equation}
   \begin{pmatrix} 0  & 1 \\ 1 & 0 \end{pmatrix}
    \begin{pmatrix}
    2 & 3x \\
    3 & 2x
   \end{pmatrix}
   \begin{pmatrix} 0  & 1 \\ 1 & 0 \end{pmatrix}
  \; = \;
    \begin{pmatrix}
    2x & 3 \\
    3x & 2
   \end{pmatrix}
  \; = \;
    x \begin{pmatrix}
    2 & 3x^{-1} \\
    3 & 2x^{-1}
   \end{pmatrix}.
\end{equation}
Hence the total number of triangles $f^{}_{n}(k)=f^{+}_{n}(k)+f^{-}_{n}(k)$ 
obeys the symmetry $f_{n}(k)=f_{n}(n-k)$ for all $0\le k\le n$.

These observations have some interesting consequences. For instance,
it can been shown that for even $n$
\[
f_n^+(0)<f_n^-(0)<f_n^+(1)<f_n^-(1)<\ldots
<f^+_n(\tfrac{n}{2}-1)<f^-_n(\tfrac{n}{2}-1)<
f^+_n(\tfrac{n}{2})=f_n^-(\tfrac{n}{2})
\]
and for odd $n$ 
\[
f_n^+(0)<f_n^-(0)<f_n^+(1)<f_n^-(1)<\ldots<f^+_n(\tfrac{n-1}{2}-1)
<f^-_n(\tfrac{n-1}{2}-1)<f^+_n(\tfrac{n-1}{2})=f_n^-(\tfrac{n+1}{2})\, .
\]

It follows that $f_{n}(k)>f_{n}(k-1)$ for $1\le k\le [\frac{n}{2}]-1$,
and the symmetry property $f_{n}(k)=f_{n}(n-k)$ then implies a 
corresponding inequality for large $k$.

\section{Diffraction of pinwheel point sets}

While it has been shown that the autocorrelation of the pinwheel point
set is circularly symmetric \cite{MPS06}, it is not known whether it
is concentrated on sharp rings only, or whether there is also a
continuous component. Direct numerical computations are of no help
here, because any finite patch only contains a small number of
independent directions, and is clearly not a good approximation of the
limit structure. However, as shown in \cite{BFG07a,BFG07b}, the
knowledge that the diffraction is circularly symmetric can be
exploited to derive an approximation of the diffraction based on a
radial version of the Poisson summation formula; see \cite{BFG07a} for
details. This results in an expansion in terms of Bessel functions. As
shown in \cite{BFG07a,BFG07b}, this results in a radial distribution
of the diffraction intensity that shows clear maxima for distances
corresponding to those of a square lattice, so the radially discrete
part resembles that of a powder diffraction of a square
lattice. However, as the pinwheel structure contains a hierarchical
sequence of scaled square lattices, the intensity distribution is not
the same as for the square lattice, but deviates in a characteristic
fashion. In addition, these numerical data provide an indication that
there is also a continuous component.

Here, we consider the slightly generalised case of diffraction from
weighted pinwheel sets. Circular symmetry also holds for the
autocorrelation of a weighted pinwheel control point set, with weights
depending on the two chiralities of the triangular tiles. This follows
from \cite[Thm.~6.1]{F08}: the triangles can be regarded as two
prototiles, a left-handed and a right-handed one, and each individual
prototile then shows circular symmetry, too.

Denoting the sets of pinwheel control points for the two chiralities
by $\varLambda_{\pm}$, with $\varLambda_{+}\cup\varLambda_{-}
=\varLambda$, we thus consider the weighted Dirac comb
\begin{equation}
  \omega \; = \; 
         \alpha_{+}\,\delta_{\varLambda_{+}}+
         \alpha_{-}\,\delta_{\varLambda_{-}}  
         \; := \; \sum_{x\in\varLambda_{+}} \alpha_{+}\,\delta_{x} +  
            \sum_{x\in\varLambda_{-}} \alpha_{-}\,\delta_{x}\, ,
\end{equation}
where $\alpha_{\pm}$ are arbitrary, in general complex, weights. In
this paper, we restrict ourselves to values
$\alpha_{\pm}\in\{-1,0,1\}$, with the case $\alpha_{+}=\alpha_{-}=1$
corresponding to the pinwheel diffraction considered in
\cite{BFG07a,BFG07b}.

The approximations to the radial density distribution shown below are
obtained by computing approximate autocorrelation coefficients
$\eta(r)$ for distances $r$, obtained from finite approximants of the
pinwheel tiling, up to a suitable maximum distance. The approximation
to the autocorrelation then has the form
\begin{equation}\label{eq:auto}
   \gamma_{\omega}
    \; = \; \sum_{r \in \mathcal{D}}
    \eta(r) \,\mu^{}_r \, ,
\end{equation}
where $\mu_{r}$ denotes the uniform measure on the unit circle, and
\begin{equation}\label{eq:D}
\mathcal{D}\subset\Bigl\{ \sqrt{ \frac{p^2+q^2}{5^{\ell}}} \;\Big| \; 
  p,q,\ell \in \mathbb{N} \Bigr\}
\end{equation}
denotes the distance set for the pinwheel tiling. Here,
the assumption of circular symmetry enters --- the approximate
autocorrelation coefficients obtained from a finite patch are used to
define a circularly symmetric autocorrelation measure.  {}From
Equation~\eqref{eq:auto}, the approximate radial distribution of the
diffraction intensity $I(k)$ is obtained as the weighted sum of the
Fourier transform of the uniform measure $\mu_{r}$. As the latter is
given by $\widehat{\mu}_{r}(k)=J_{0}(2\pi |k| r)$, this results in
finite sums of Bessel functions $J_{0}$ which show strong
oscillations; compare \cite{BFG07a,BFG07b} for details. We note in
passing that some values for frequencies of distances are known
exactly, and they can in principle be computed from the inflation
rule; see \cite{M09}.

\begin{figure}[t]
\centerline{\includegraphics[width=0.8\textwidth]{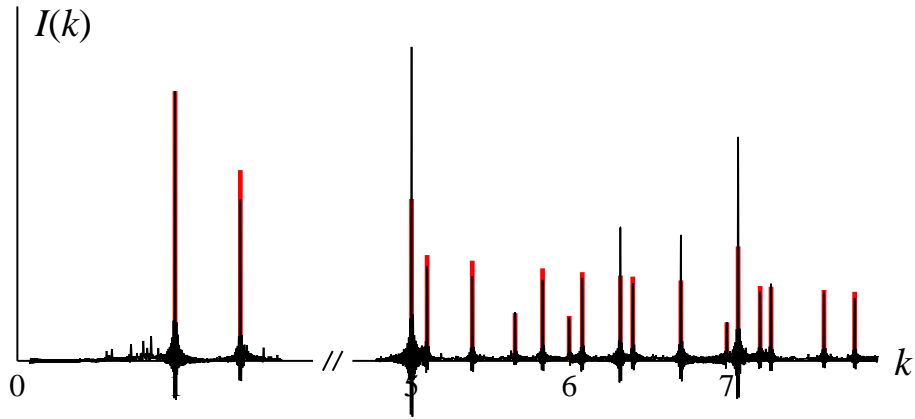}\vspace*{-2ex}}
\caption{\label{fig:sqapprox} Part of the approximated radial intensity
  distribution for the pinwheel diffraction
  ($\alpha_{+}=\alpha_{-}=1$), compared with the exact diffraction for
  a square lattice powder (red bars).}
\end{figure}

\subsection{Systematics of peak intensities}

Figure~\ref{fig:sqapprox} shows the numerical approximation for the
diffraction of $\delta_{\varLambda}$, which is the pinwheel point sets
with weights $\alpha_{+}=\alpha_{-}=1$. The corresponding result for a
square lattice powder is also shown, with relative normalisation
chosen to make the first peak match. Clearly, as observed in
\cite{BFG07a,BFG07b}, the main peaks in the intensity are well
reproduced, but there are systematic deviations in the peak
intensities. The latter can qualitatively be explained as arising from
distances in $\mathcal{D}$ of Equation~\eqref{eq:D} with denominators
$5^\ell$ with $\ell>0$.

\begin{figure}[t]
\centerline{\includegraphics[width=0.45\textwidth]{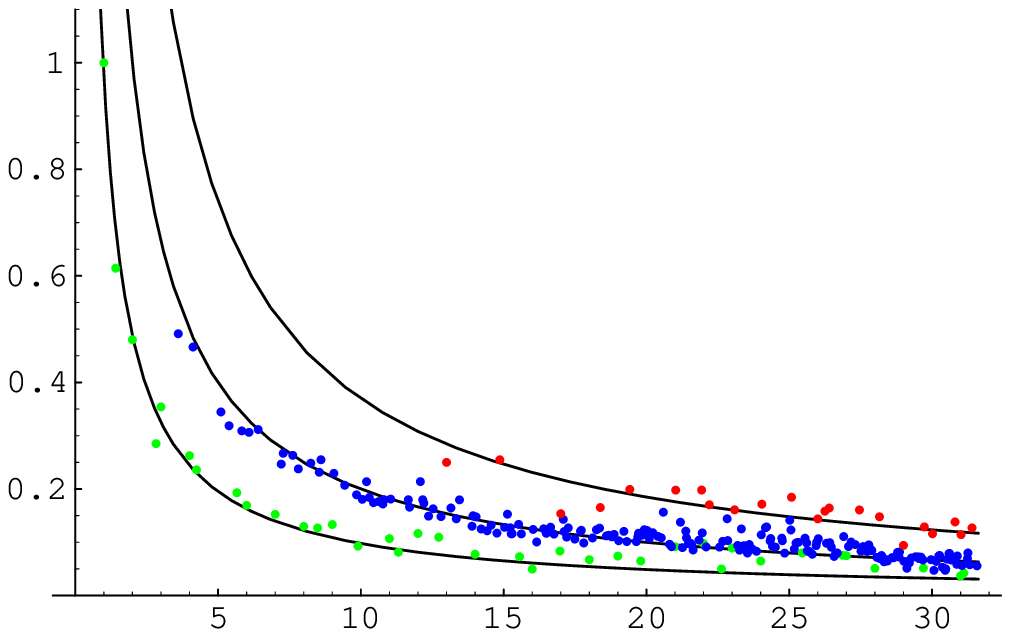}
\includegraphics[width=0.45\textwidth]{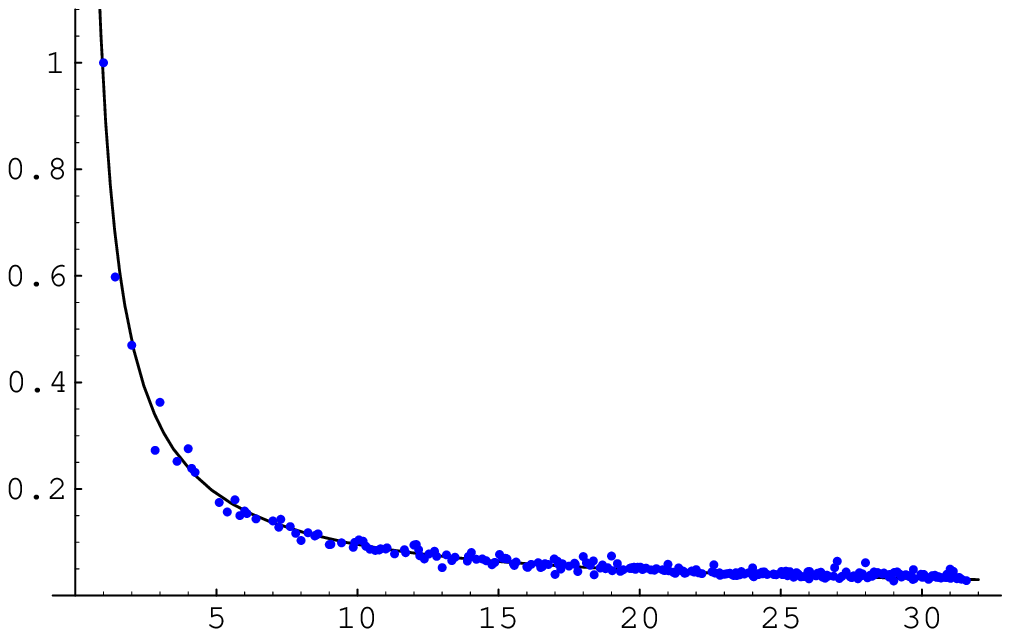}\vspace*{-1.5ex}}
\caption{\label{fig:ratplot1} Intensity ratios $I(k)/I(1)$ of peaks
  for values of $k$ such that $k^2$ is not divisible by $5$, as a
  function of $k$. On the left, different colours refer to different
  number of prime factors equal to $1$ modulo $4$ (either $0$ (green),
  $1$ (blue) or $2$ (red)). Note that equal factors are counted here;
  for instance, the lowest $k$ in the second group is $k=\sqrt{13}$,
  and in the third group it is $k=\sqrt{169}=13$.  On the right, these
  data have been collapsed as described in the text. Lines are least
  square fits to functions of the form $c/k$ with constant $c$.}
\end{figure}

To get a better picture of the peak intensities, we consider
separately the integrated intensities on rings at values of $k$
according to the highest power of $5$ that divides $k^{2}$ (which will
be $0$, $1$ or $2$ in the examples below).  Figure~\ref{fig:ratplot1}
shows the result for the case where $k$ is not divisible by $5$. On
the left, you can clearly distinguish three groups of data, indicated
by different colours, and for each group of data the decay of
intensities with increasing $k$ is well approximated by a function of
the type $c/k$, where $c$ is a constant.

\begin{figure}[t]
\centerline{\includegraphics[width=0.45\textwidth]{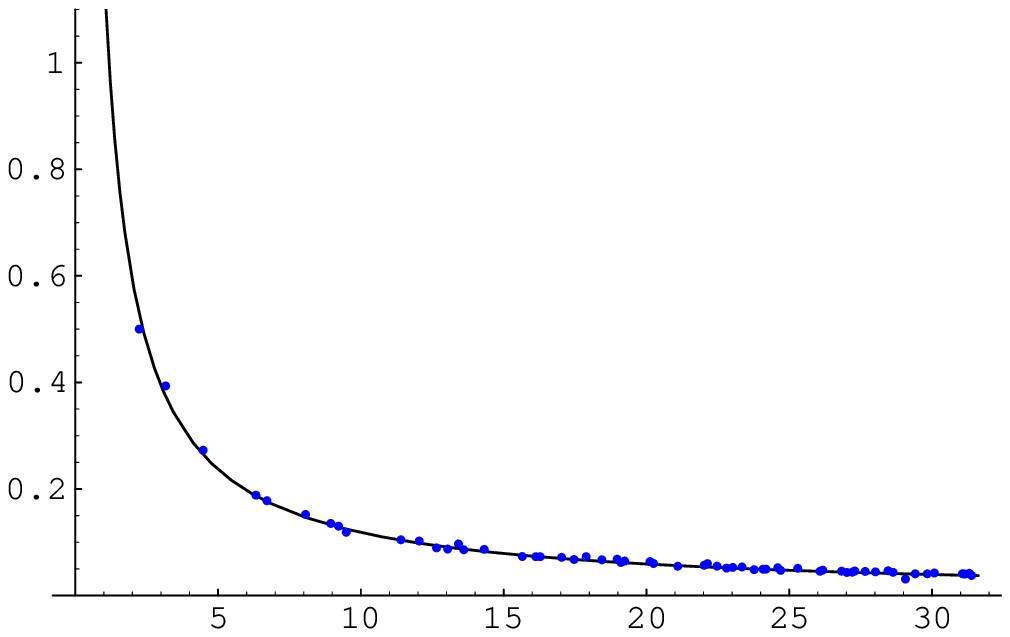}
\includegraphics[width=0.45\textwidth]{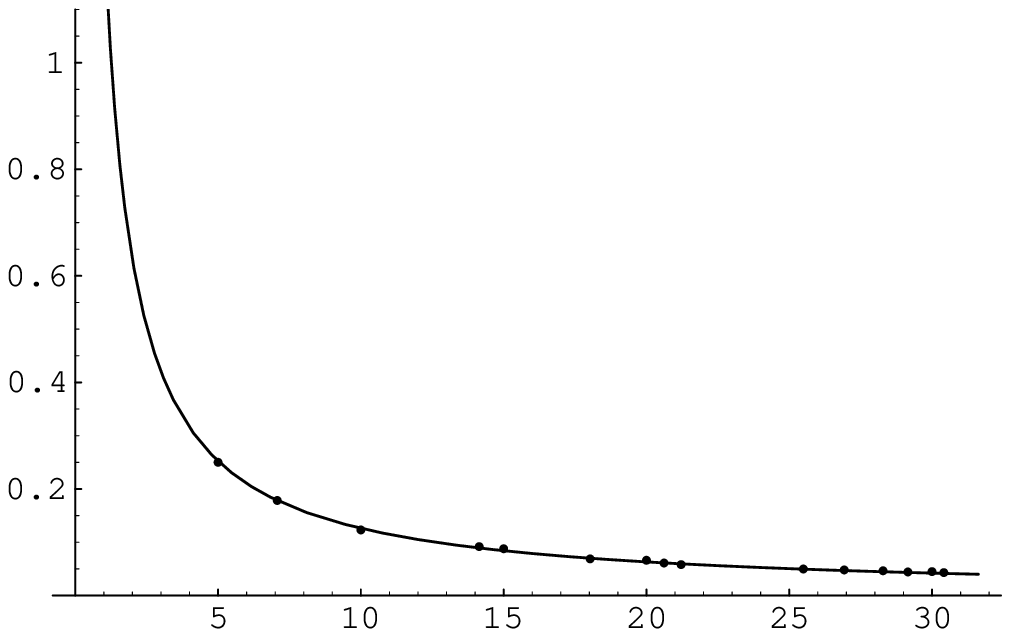}\vspace*{-1.5ex}}
\caption{\label{fig:ratplot2} Same as right part of Figure~\ref{fig:ratplot1},
for values of $k$ such that $k^2$ is divisible by $5$ but not $5^2$ (left) and
such that $k^{2}$ is divisible by $5^2$ but not by $5^3$ (right).}
\end{figure}

The three groups of data are distinguished by the number of prime
factors of $k^2$ that are equal to $1$ modulo $4$. In fact, dividing
the intensity ratios by $2^s$, where $s$ is the sum of powers of all
prime factors of this type, collapses the data set onto a single
curve, as shown in the right part of Figure~\ref{fig:ratplot1}.  This
observation is in line with the similarity to the square lattice
powder diffraction seen in \cite{BFG07a,BFG07b}.

The corresponding collapsed data for peaks at values of $k$ which are
multiples of $5$ and $5^2$ are shown on the left and right of
Figure~\ref{fig:ratplot2}, respectively. Again, the observed decay of
the intensities is inversely proportional to $k$. This indicates that
our qualitative understanding of the peak intensities is correct,
though it is not clear whether this also holds quantitatively for the
diffraction of an infinite pinwheel tiling point set.

\subsection{Diffraction of weighted pinwheel point sets}

\begin{figure}[t]
\centerline{\includegraphics[width=0.8\textwidth]{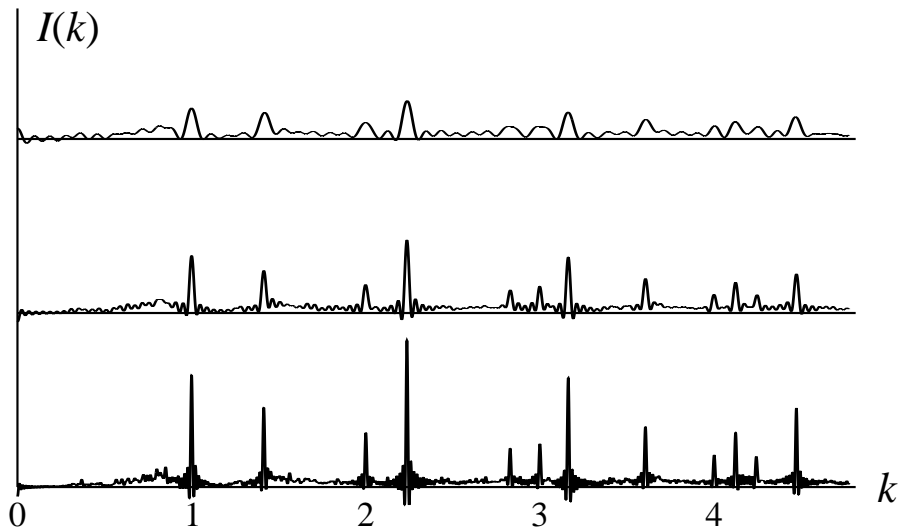}\vspace*{-2ex}}
\caption{\label{fig:pinint} Three approximations of the radial intensity
distribution of the diffraction of the Dirac comb $\delta_{\varLambda_{+}}$.}
\end{figure}

In order to get a better indication of the possible continuous
contribution of the pinwheel diffraction, it is advantageous to
consider weighted pinwheel point sets. As discussed above, 
the resulting autocorrelation is still circularly symmetric, and hence
the same approach as outlined above can be applied to obtain
approximations of the radial intensity
distribution. Figure~\ref{fig:pinint} shows the result for the case
$\omega=\delta_{\varLambda_{+}}$ ($\alpha_{+}=1$ and $\alpha_{-}=0$),
where only the half of the pinwheel point set is considered.  The
three spectra correspond to an increasing number of terms in the
approximation, and increasing accuracy of the estimated correlation
coefficients from larger patches of the pinwheel tiling constructed by
successive inflation. The result very closely resembles that for the
complete pinwheel point set, and may well be identical apart from the
overall intensity normalisation due to the different density of
scatterers.

\begin{figure}[t]
\centerline{\includegraphics[width=0.8\textwidth]{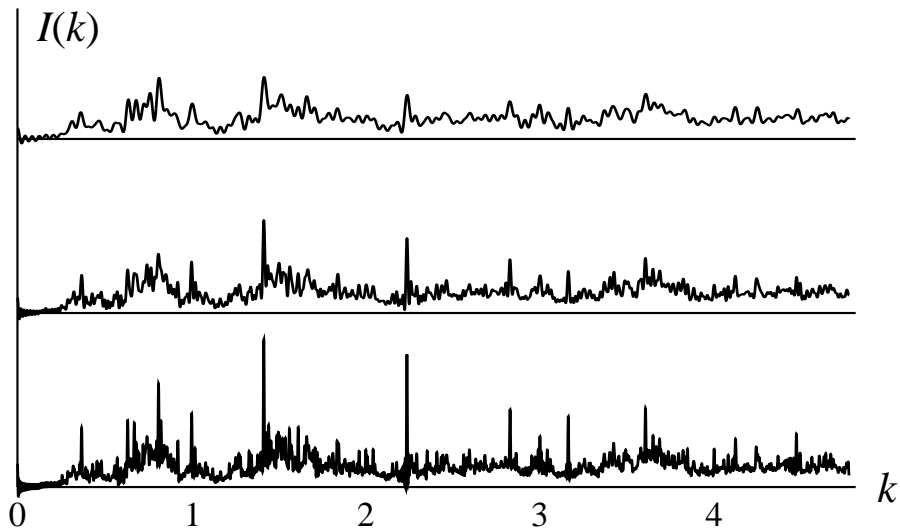}\vspace*{-2ex}}
\caption{\label{fig:pinsign} Three approximations of the radial
  intensity distribution of the diffraction of the Dirac comb $\omega$
  for the balanced case with $\alpha_{+}=1$ and $\alpha_{-}=-1$.}
\end{figure}

More interesting is the case of $\alpha_{+}=1$ and
$\alpha_{-}=-1$. Since this is a balanced case, the average scattering
strength is zero, and there is no peak at $k=0$. Therefore, a continuous 
background may be easier to spot in this case. Figure~\ref{fig:pinsign}
shows three increasingly accurate approximations of the 
radial diffraction intensity for this case.

There is not only an indication of the presence of a continuous
background, similar to what was found in \cite{BFG07b}, but also a
number of new features appear to develop (in particular for small
values of $k$), which may correspond to additional peaks for the
infinite system. However, the approximation may not yet be good enough
to draw firm conclusions.

\section{Summary}

The diffraction of substitution tilings with continuous rotational
symmetry still remains to be completely understood. In this paper, we
presented some observations on combinatorial properties of the
pinwheel tiling and on its diffraction measure, using the
approximation introduced in \cite{BFG07a,BFG07b}. {}From the numerical
data, we corroborate the conclusions of \cite{BFG07a,BFG07b} on the
similarities between the pinwheel diffraction and a square lattice
powder.  Considering the balanced case of zero average scattering
strength provides an indication that there might not only be a
continuous contribution as conjectured in \cite{BFG07b}, but that
there might also be additional sharp rings with finite intensity. This
observation warrants further investigation. It would also be
interesting to compare this result with the diffraction of other
tilings, such as the generalised pinwheel tilings introduced by Sadun
\cite{S98} or the tipi tilings in \cite{F08}.

\section*{Acknowledgements} 
The authors thank Michael Baake, Dirk Frettl\"{o}h and Ha\"{\i}ja
Moustafa for useful discussions. XD is grateful to the Department of
Mathematics and Statistics at the Open University for the hospitality
during an extended visit in 2009, where some of this work was
completed.

\end{document}